\documentclass[journal]{IEEEtran}

\usepackage{titlesec}

\ifCLASSINFOpdf
\else
\fi

\usepackage{url}
\usepackage{verbatim} 
\usepackage{amsmath,amssymb,amsfonts}
\usepackage{algorithmic}
\usepackage{graphicx}
\usepackage{textcomp}
\usepackage{xcolor}
 
\usepackage{tabularx}
\usepackage{graphicx}
\usepackage{adjustbox}
\usepackage{multirow}
\usepackage{caption}
\usepackage{colortbl}
\usepackage[numbers,sort&compress]{natbib}

\begin{document}
%
\title{Adapting LoRaWAN to the Open-RAN Architecture}
%
%
%

\author{Sobhi Alfayoumi, 
        Joan Meli\`a-Segu\'i,
        Xavier Vilajosana,

\thanks{S. Alfayoumi and X. Vilajosana were with the Wireless Networks Research Lab, Universitat Oberta de Catalunya and Worldsensing S.L, Barcelona, Spain}
\thanks{J. Meli\`a-Segu\'i was with the Wireless Networks Research Lab, Universitat Oberta de Catalunya, Barcelona, Spain}

}

\maketitle

\begin{abstract}
This article proposes O-LoRaWAN, an adaptation of the LoRaWAN architecture into a modular network architecture based on the Open RAN (O-RAN) principles. In our vision, standardization of the network components and interfaces will enable the reuse of network functions, and thus, foster an accelerated tailoring of the network functions to the changing application demands. 
LoRaWAN shares similarities to cellular networks and becomes an interesting candidate for a transformation to the O-RAN standard. 
In the article we draw several transition strategies, these include the reorganization of the LoRa gateway functions into Radio and Distributed Units; enhancing network performance with RAN Intelligent Controllers exploiting the network data; and the standardization of the management and orchestration of network components. 
Key for that adaptation are the O-RAN interfaces. Along the article, we analyze them and suggest protocol extensions or adjustments for compatibility and interoperability between network components, advocating for the design of extensible protocols. 

\end{abstract}



\section{\textbf{Introduction}}
\label{sec:introduction}

The growth in wireless technology demands network architectures that are not only efficient and flexible but also scalable. This requirement is propelled by the need for dynamic, adaptable, and high-performance communication systems in an evolving digital landscape. Our research introduces a significant approach to redefining wireless communication networks by incorporating Open Radio Access Network (O-RAN) principles within LoRaWAN systems. This initiative is aimed at augmenting network performance, enhancing system agility, and embedding Artificial Intelligence (AI) driven management capabilities to foster smarter, more responsive wireless networks. Unlike merely merging different technologies, this strategic adoption of O-RAN methodologies seeks to substantially standardize LoRaWAN networks, transforming them into more flexible, scalable, and interoperable frameworks.
This approach is part of a broader vision to develop a unified, flexible, and versatile network framework that spans across various wireless systems, starting by cellular, but growing to other systems like LoRaWAN and possibly WiFi. It sets new benchmarks for network architectures by leveraging shared resources and modular designs, paving the way for a smarter and more interconnected wireless communications industry.



Along the article we introduce the following key contributions:\\ 
\textbf{Conceptualization of a broader O-RAN concept:} by identifying the adaptation layers, fostering a support for diverse technologies, finding the functional barriers and needs for a cross-technology scalable framework.\\
\textbf{Integration of LoRaWAN into the O-RAN Architecture:} by adapting the LoRaWAN network building blocks to the O-RAN principles. Proposing a functional split for the legacy LoRaWAN network component to align with the O-RAN principles.\\
\textbf{Posing the basis for a AI-augmented LoRaWAN infrastructure:} by advancing  the LoRaWAN framework towards a standardized and scalable approach to introduce AI-driven optimization.\\
\textbf{Protocol integration and compatibility:} by identifying and proposing key adaptations to maximize the O-RAN framework reuse while raising indications to the O-RAN designers to ensure the framework is adaptable to various network technologies.

\section{\textbf{LoRaWAN Overview}}
\label{sec:LoRaWAN}
\begin{figure}[ht]
\centering
\includegraphics[width=3.4in]{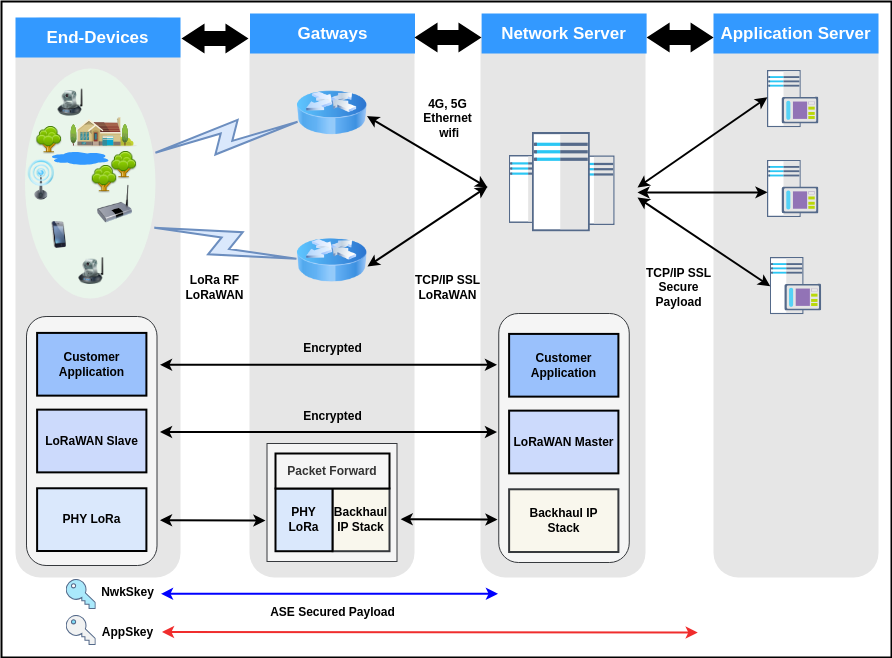}
\caption{Typical LoRaWAN network architecture.}
\label{fig:lorawan}
\end{figure} 

LoRaWAN is a consolidated Low Power Wide Area Network (LPWAN) standard developed by the LoRa Alliance. It is characterized by its energy-efficient operation, limited data transmission rates, and kilometre-scale communication range. This technology was developed by Cycleo, a company that was acquired by Semtech Corporation in 2010. 
LoRaWAN supported the development of the outdoor and large scale IoT industry, thanks to addressing key limitations of previous networking technologies, especially robustness, low power operation and communication range ~\cite{adelantado2017understanding,9422331}. In addition, it also enabled ad-hoc and self-operated networks deployments, which enabled certain industries to address use cases in remote and complex areas. 

The layout of the legacy LoRaWAN network's deployment is shown in Fig.\ref{fig:lorawan}. A typical LoRaWAN network architecture builds on a star-of-stars topology, and it includes the following key components:\\
\textbf{End Devices:} autonomous sensors and IoT devices using LoRa RF Modulation for secure wireless connectivity. They generate digital environmental data for various IoT applications.\\
\textbf{Gateways:} act as communication links between the end devices and the LoRaWAN Network Server (LNS) which helps in the efficient data transfer with various backhaul options such as Wi-Fi, Ethernet, Cellular, and Satellite. These gateways serve as relay points on which RF messages from numerous end-devices are validated and transmitted, while offering some processing capability for handling message integrity and forwarding metadata to the LNS where the best appropriate signal for a response will be determined. It is capable of handling multi-channel operations and do not need device pairing to provide wide network coverage and flexibility.\\
\textbf{Network Server:} The LNS is the hub of the LoRaWAN infrastructure and is essential for network management and security assurance. It ensures the security of the managed data through 128-bit AES encryption, authenticates devices, and controls network traffic accordingly. Primary operations comprise the management of device addresses, the message encryption and authentication, alert to non-confirmations and network adaptation via dynamic Data Rate adaptation through the Adaptive Data Rate (ADR) protocol. The LNS manages the origination and termination of uplink and downlink communications, properly routing traffic. It also manages message redundancy by merging the messages from the different gateways into one.\\
\textbf{Application Server:} has the role of securely processing, managing or forwarding uplink and downlink application data. Security-wise manages the application keys and is capable of decoding the application layer frame payload, conforming the two layer security architecture. Multiple Application servers can interact with a LNS, enabling separation of applications across the LoRaWAN network infrastructure. 

\section{\textbf{Open-RAN Overview}}
\label{sec:Open-RAN} 

O-RAN marks a significant shift in how radio access networks are built, moving from traditional, closed systems to a more open, flexible, and interoperable setup. This change is driven by the need for networks that are more adaptable and extensible and more cost-effective, capable of handling the increasing demands for data and connectivity. By using a vendor-neutral approach, O-RAN allows different network parts from various suppliers to work together smoothly, overcoming the limitations of older systems that were locked into single vendors and their proprietary technologies.
At its heart, O-RAN is built on three main principles:
Openness, implementing an open architecture, promoted by an independent alliance, with well-known and accessible interfaces and components. Intelligent, by enabling the integration of AI methods as part of the network functionality, for example for predictive operations or security. Interoperability, built on standards that allow equipment or software components from many different vendors to be interconnected.
 
The O-RAN Alliance, formed in 2018, supports these ideas by focusing on the organization of the base station functions (the 3GPP NR 7.2 split) \cite{3GPP2017}. This aims to improve how base stations work and communicate through smart controllers and open interfaces, aligning with the aforementioned principles.

\begin{figure}[t]
\centering
\includegraphics[width=3.4in]{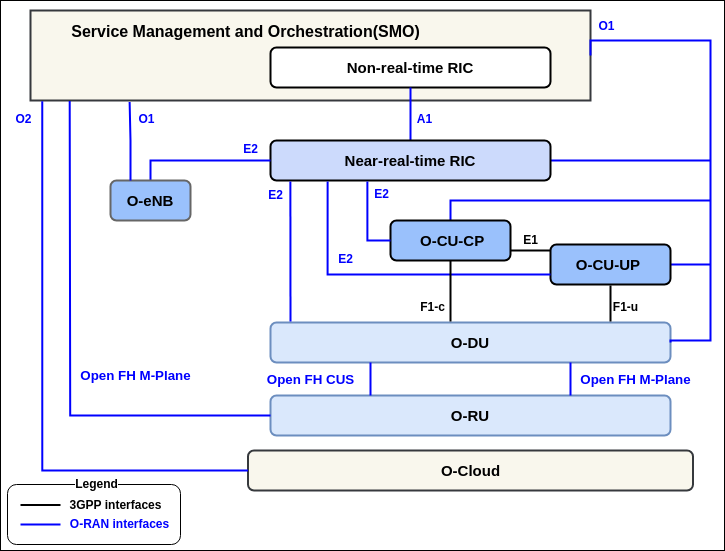}
\caption{Logical Architecture of O-RAN \cite{O_RAN_WG1_2023}.}
\label{fig:openran}
\vspace{-3mm} 
\end{figure} 

\subsection{\textbf{O-RAN Architecture}}
The O-RAN architecture is depicted in Fig.\ref{fig:openran}. In what follows its main components are summarized. For further details on O-RAN architecture, we refer to O-RAN specifications and comprehensive surveys\cite{O_RAN_WG1_2023,O_RAN_WG2_NONRTRIC_ARCH_2023,O_RAN_WG2_A1GAP_2023,O_RAN_WG2_NONRTRIC_FUNARCH_2021,O_RAN_WG3_RICARCH_2023,O_RAN_WG4_2023,O_RAN_WG6_CADS_2023, O_RAN_WG6_O2GAP_2022, O_RAN_WG3_E2GAP_2023, O_RAN_WG1_O1_2021, polese2023understanding}.\\
\textbf{Near-Real-Time RAN Intelligent Controller (Near-RT RIC):} Positioned at the network edge and it is designed to manage control loops within a timeframe of 10 milliseconds to 1 second  \cite{O_RAN_WG3_RICARCH_2023}. It interfaces with Distributed Units (DUs), Central Units (CUs), and legacy O-RAN-compliant LTE evolved Node Bases (eNBs), overseeing QoS for numerous User Equipments (UEs). It utilizes the concept of xApps, micro-applications connected to the O-RAN database or interfaces, for RAN data processing and control command issuance through the E2 interface, with the goal to enhance network intelligence and efficiency.\\ 
\textbf{Service Management and Orchestration (SMO):} Central management unit for orchestrating 5G services and elements, and merging functionalities within the O-RAN system, including O-Cloud, Near-RT RIC, and Non-RT RIC. It enforces policies, manages resources, and ensures optimal network performance \cite{O_RAN_WG2_NONRTRIC_ARCH_2023}\cite{O_RAN_WG2_NONRTRIC_FUNARCH_2021}.\\
\textbf{Non-Real-Time RAN Intelligent Controller (Non-RT RIC):} Operating on a timescale of several seconds to minutes, the Non-RT RIC focuses on network optimization and policy-driven decisions \cite{O_RAN_WG2_NONRTRIC_ARCH_2023}\cite{O_RAN_WG2_NONRTRIC_FUNARCH_2021}. It facilitates intent-based management and intelligent orchestration, interfacing with the Near-RT RIC via the A1 interface, and with network elements through O1 and O2 interfaces for comprehensive network management and xApps/rApps deployment.\\
\textbf{O-RAN Central Unit (O-CU): }The O-CU manages high-level radio network functions and protocols, divided into O-CU Control Plane (O-CU-CP) for signaling and communications, and O-CU User Plane (O-CU-UP) for data traffic management. It interfaces with RICs and the SMO, supporting E2 and O1 interfaces for dynamic RAN optimization and comprehensive network management.\\
\textbf{O-RAN Distributed Unit (O-DU):} Essential for real-time RAN functions, the O-DU ensures efficient connection with RUs and CUs, managing lower-layer processes crucial for data transmission and radio resource management. It interacts with RICs for RAN control and optimization, and with SMO for network management, embodying a core element for 5G network performance. This robust design supports network slicing and QoS management, aligning with O-RAN's goals for a flexible and open network ecosystem \cite{3GPP2017}.\\
\textbf{O-RAN Radio Unit (O-RU):} Managing RF functions and Low-PHY processing, the O-RU is critical for quality radio transmission and reception. It supports a flexible split 7-2 \cite{3GPP2017}, for precoding functions, enhancing network adaptability. Additionally, it interfaces with SMO through Fronthaul M-Plane and O1 interface, contributing to O-RAN's open and flexible network vision.\\
\textbf{O-Cloud:} Cloud-based platform that supports the deployment and management of O-RAN components, offering scalability, flexibility, and enhanced performance. Managed by SMO, its goal is to enhance network efficiency and scalability, aligning with O-RAN's vision for an open and standardized network environment \cite{O_RAN_WG6_CADS_2023}. 
 

\subsection{\textbf{O-RAN Interfaces}}
In the O-RAN architecture, well-defined interfaces are key to connecting network components and enabling openness, extensibility and interoperability in 5G systems. The following interfaces serve as communication links between the components contributing to the \textit{separation of concerns} in the network architecture.
Open Fronthaul (FH) Interface specified in \cite{O_RAN_WG4_2023}, E2 Interface specified in \cite{O_RAN_WG3_E2GAP_2023}, O1 Interface specified in \cite{O_RAN_WG1_O1_2021}, A1 Interface specified in \cite{O_RAN_WG2_A1GAP_2023}, and O2 Interface specified in \cite{O_RAN_WG6_O2GAP_2022}. These interfaces enable the communication of the network components and are the key element to support the integration of other network technologies. As long as the network components meet the interface, component reuse and network functionalities could be developed as micro-applications. 






\section{\textbf{O-LoRaWAN architecture}}
\label{sec:O-LoRaWAN_architecture}

\begin{figure}[h]
\centering
\includegraphics[width=3.4in]{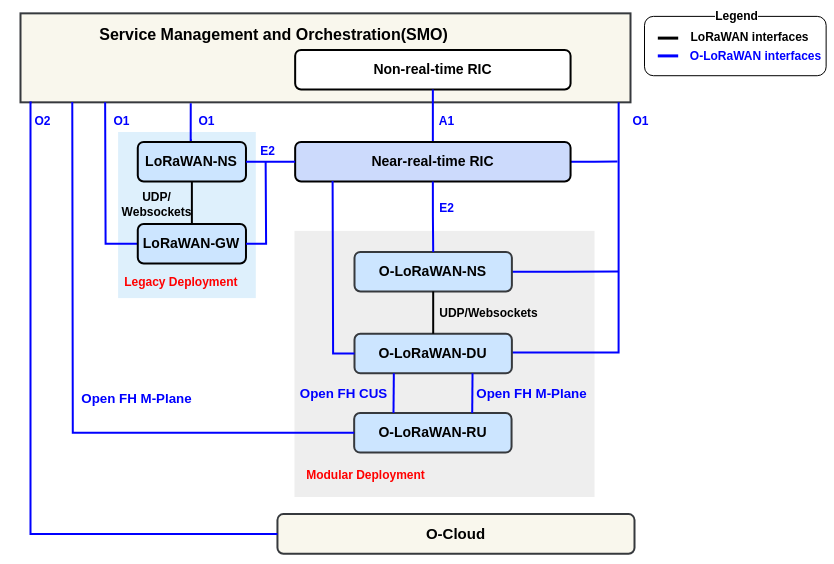}
\caption{Proposed Framework for Strategic Adaptation of LoRaWAN Architecture to the  O-RAN.}
\label{fig:o_lorawan}
\vspace{-5mm}  
\end{figure}

In this section, we introduce the O-LoRaWAN architecture, illustrated in Fig.~\ref{fig:o_lorawan}. O-LoRaWAN adapts the LoRaWAN architecture to the O-RAN principles. It provides two operational model deployments: the \textit{Legacy Deployment} and the \textit{Modular Deployment}. In what follows, we will delve into the innovative features of the Modular Deployment, noting that the traditional Legacy Deployment is detailed in section \ref{sec:LoRaWAN} and is well-documented in  \cite{adelantado2017understanding, 9422331, LoRaWAN2017Specification}.

\begin{figure}[!h]
\centering
\includegraphics[width=3.4in]{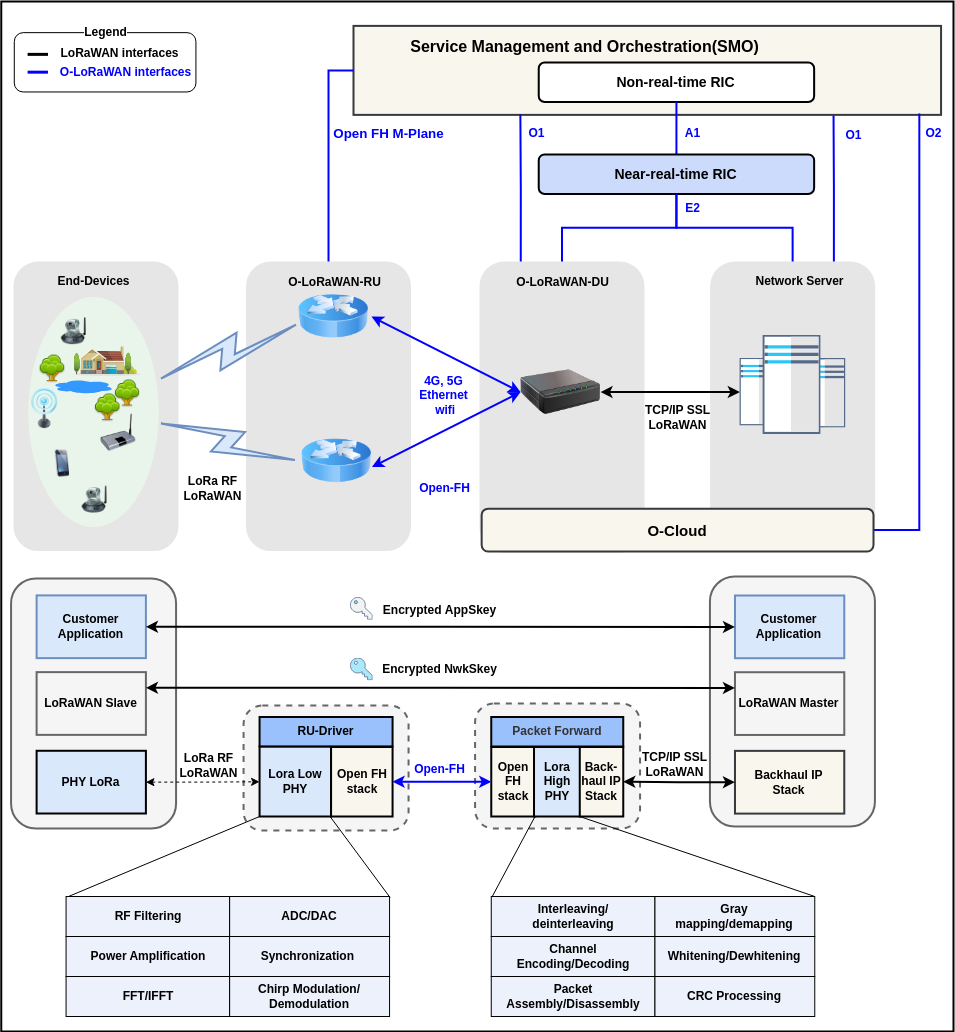}
\caption{O-LoRaWAN Architecture: Protocol Stack Distribution and Gateway Functional Split between O-LoRaWAN-RU and O-LoRaWAN-DU.}
\label{fig:O_lorawan_protocol_stack}
\vspace{-6mm}
\end{figure}

The Modular Deployment redefines LoRaWAN gateway functions, following the O-RAN's 7-2 approach, with a clear division between the O-LoRaWAN-DU, which manages signal processing and data management, and the O-LoRaWAN-RU, responsible for signal transmission and reception as seen in Fig.\ref{fig:O_lorawan_protocol_stack}. This approach reorganizes the LoRaWAN architecture functions in standardized components, enabling a wider component reuse and fostering a micro-service component development across RANs.
Next, we explore how the responsibilities of the classic gateway are allocated to the newly introduced modules.


\subsection{\textbf{O-LoRaWAN-DU (Distributed Unit)}}
The O-LoRaWAN-DU provides flexibility and scalability to the O-LoRaWAN framework by adopting a split architecture, separating higher-level data processing from direct radio signal handling. This approach allows for a logical separation of network components facilitating component upgrades or even enabling the processing of frames from multiple radio units for performance optimization or scalability. Furthermore, network functionalities can be developed independently considering the interfaces at each of the components and as functionalities are systematically split, even for different RANs, functionality reuse can be achieved across RANs.  

For a detailed view of its uplink (UL) and downlink (DL) responsibilities, see Table.\ref{tab:comprehensive_uplink_downlink_responsibilities}.

\subsection{\textbf{O-LoRaWAN-RU (Radio Unit)}}
The O-LoRaWAN-RU serves as the radio interface in the LoRaWAN architecture, handling essential radio communication tasks with User Equipment. Its core responsibilities include processing radio frequency signals and managing the initial stages of signal conversion. The RU enables a separation of the legacy gateway functionalities, enabling as in the case of cellular gNBs, the functional splitting that facilitates network scaling, adaptation to real time operation or specific radio customization. 
In what follows a brief overview of its fundamental roles is presented:\\
\textbf{RF Reception/Transmission:} Captures and transmits RF signals to and from end-devices.\\
\textbf{Analog-to-Digital (ADC)/Digital-to-Analog Conversion (DAC):} Converts analog signals to digital IQ samples for uplink and reverses the process for downlink.\\
\textbf{Chirp Processing and Frequency Domain Processing:} Handles chirp modulation and demodulation, which is characteristic of LoRa technology. Utilizes FFT and IFFT to translate signals between the time and frequency domains, and to prepare IQ samples for further processing.\\
\textbf{Initial Signal Processing:} Improves signal quality through amplification and filtering, including low-noise and power amplification, as well as band-pass filtering.\\
\textbf{Timing and Frequency Synchronization:} Ensures signal timing and frequency are aligned with network protocols for accurate demodulation and decoding.\\
\textbf{Performance Reporting:} Collects and reports on transmission efficacy and operational metrics.\\
\textbf{RU-Driver:} Manages the software interface between the RU's physical layer components and the DU.\\



\begin{table}[th]
\renewcommand{\arraystretch}{2}
\begin{center}
\begin{adjustbox}{width=.5\textwidth}
\begin{tabular}{ p{1.5cm} p{5.25cm} p{5.25cm} }

\hline
\rowcolor{blue!15}
\textbf{Functionality} & \textbf{UL Responsibilities} & \textbf{DL Responsibilities} \\

\hline
Advanced Modulation/ Demodulation  & Performs sophisticated demodulation of IQ samples to retrieve uplink data. Includes enhancements for signal clarity and noise suppression. & Executes complex modulation on IQ samples for transmission, optimizing signal strength and clarity for diverse network environments. \\

Channel Decoding/ Encoding & Implements error correction algorithms, such as Hamming and Reed-Solomon codes, to preserve data integrity and facilitate reliable communication. & Applies robust encoding methods to data, ensuring transmissions are resilient to errors and can be correctly decoded by end devices. \\

Interleaving/ Deinterleaving & Rearranges the data bits to protect against burst errors, enhancing the robustness of signal reception. & Sequences data bits to prevent burst error impacts on downlink transmissions, ensuring reliable data delivery. \\

Whitening/ Dewhitening & Applies data randomization to maintain a balanced distribution of bits, preventing pattern-related transmission issues. & Restores the original data order upon reception, reversing the whitening process to maintain data integrity. \\

CRC Processing & Validates the integrity of incoming data by checking CRC values, flagging errors for retransmission or correction. & Appends CRC values to outgoing data packets, facilitating error detection at the receiving end. \\

Packet Assembly/ Disassembly & Deconstructs incoming packets into actionable data, while also handling data routing to appropriate network layers. & Assembles data into structured packets, complete with headers and payloads, ready for transmission to end devices. \\

Adaptive Data Rate (ADR) Management & Dynamically adjusts the data rate and power based on signal quality and network conditions, optimizing uplink communications. & Tailors transmission power and data rate for downlink, ensuring efficient use of network resources and maintaining connection quality. \\

MAC Layer Processing & Manages Medium Access Control functions, enforcing channel access rules and synchronization of uplink transmissions. & Oversees MAC functions for downlink, including scheduling transmissions and managing duty cycles to comply with regulations. \\

Network Layer Functions & Orchestrates routing, addressing, and data forwarding within the network, ensuring efficient and logical data flows. & Manages network layer protocols for downlink, facilitating efficient packet delivery and maintaining network organization. \\

Security Protocol Integration & Integrates and manages security protocols, ensuring that uplink data is securely encrypted and authenticated. & Prepares downlink data packets with the required security measures, ensuring end-to-end encryption and data protection. \\

Timing and Synchronization & Maintains precise timing for uplink processes, ensuring synchronization with network server operations. & Coordinates downlink timing to align transmissions with the network's time slots and synchronization standards. \\
 
\hline
\end{tabular}
 \end{adjustbox}
\end{center}
\caption{Comprehensive Responsibilities of the O-LoRaWAN-DU in UL and DL Communications.}
\vspace{-5mm}
\label{tab:comprehensive_uplink_downlink_responsibilities}
\end{table}

\section{\textbf{RAN Intelligent Controller in LoRaWAN}}
\label{sec:RIC}
The O-RAN specification introduces the concept of RIC, enabling the deployment of network controllers or applications that build on the data or metadata transported by the network. The RIC becomes an important support for AI driven controllers that can serve for diverse purposes. Some of the control logic could operate on a very short timescale to provide rapid, real-time responses to network conditions, this is referred to as the Near-RT RIC. The Non-RT RIC extends its focus to longer-term actuation, optimizing the network and making decisions based on established policies over longer timescales (e.g seconds or minutes).
Thanks to the O-RAN modularity key RIC controllers in cellular networks can be exported to LoRaWAN while some others can be tailored to the particular nature of LoRaWAN. 

The Near-RT RIC uses special applications called xApps to tackle O-LoRaWAN's specific needs. Table~\ref{tab:xApps_rApps_functions}, presents potential xApps that could be deployed in O-LoRaWAN, illustrating their role in network optimization. These applications can be used to optimize energy consumption, data rates, and ensure regulatory compliance, making the network more adaptable. 

The Non-RT RIC focuses on strategic management and long-term planning. It uses rApps to manage network operations, develop policies, and leverage analytics for network optimization. These rApps can be designed to address the specific needs of O-LoRaWAN (as presented in Table \ref{tab:xApps_rApps_functions}), such as O-LoRaWAN modules management, device integration, specific data management and handling, security, contribute to strategic planning for network capacity, growth, and maintenance. 
\begin{table}[!h]
\renewcommand{\arraystretch}{1.7}
\begin{center}
\begin{adjustbox}{width=0.5\textwidth}
\begin{tabular}{p{5cm} p{5cm}}

\hline
\rowcolor{blue!15}
\textbf{xApps for LoRaWAN Networks} & \textbf{rApps for LoRaWAN Networks} \\[4pt]
\hline

\textbf{Energy Consumption Analysis :} Forecasts device power usage, tailoring communication strategies to enhance battery longevity. &
\textbf{Energy Efficiency:} Analyzes and optimizes energy consumption patterns to extend battery life of LoRaWAN end-nodes. \\

\textbf{Spreading Factor Adjustment:} Fine-tunes spreading factors in response to distance and signal quality. &
\textbf{Coverage Analysis:} Produces network coverage heatmaps and recommends infrastructural adjustments for improved signal strength. \\

\textbf{Regulatory Compliance Management:} Confirms devices comply with local transmission regulations, optimizing network traffic. &
\textbf{Capacity Planning:} Estimates future network load and advises on infrastructure scaling or upgrades. \\

\textbf{Optimal Gateway Selection:} Directs devices to the most suitable gateway, enhancing signal quality and network reliability. &
\textbf{Spectrum Optimization:} Manages the radio spectrum to reduce interference and maximize frequency usage efficiency. \\

\textbf{Signal Interference Resolution:} Detects and minimizes channel disruptions through proactive measures. &
\textbf{Compliance Management:} Ensures network compliance with regulatory standards and adjusts policies as necessary. \\

\textbf{Data Rate Management:} Modulates data rates to balance throughput with efficiency, adapting to network conditions. &
\textbf{Device Health Monitoring:} Identifies hardware malfunctions or wear, prompting maintenance or replacement. \\

\textbf{Transmission Quality Enhancement:} Adjusts transmission parameters to cut down packet errors. &
\textbf{Data Traffic Management:} Manages data flow within the network to prevent bottlenecks and prioritize critical communications. \\

\textbf{Battery Lifetime Estimation:} Provides estimates of end-node battery lifespans, aiding in maintenance planning. &
\textbf{Adaptive Modulation Scheme:} Selects optimal modulation schemes based on network conditions to enhance connectivity. \\

\textbf{Device Health Monitoring:} Alerts to possible hardware failures or security breaches, ensuring device reliability. &
\textbf{Maintenance and Upgrade Scheduler:} Coordinates network component maintenance and software updates to minimize disruptions. \\

\textbf{Traffic Prioritization for Critical Applications:} Ensures essential application traffic is given precedence. &
\textbf{Security Incident and Event Management:} Identifies and reacts to security incidents, strengthening network security. \\
\hline

\end{tabular}
 \end{adjustbox}
\end{center}
\caption{Functions of xApps and rApps in LoRaWAN Networks}
\label{tab:xApps_rApps_functions}
\vspace{-5mm}
\end{table}



\section{\textbf{Interfaces}}
\label{sec:Interfaces}

In this section, we conduct a careful analysis to understand the required and necessary adjustments for integrating O-RAN interfaces into an O-LoRaWAN network architecture. This examination will reveal whether the standardized O-RAN interfaces can be applied to O-LoRaWAN network structures, or whether modifications are required. Instead of developing a full-scale solution, the emphasis will be on assessing the potential for such integration. 
\begin{figure}[ht]
\centering
\includegraphics[width=3.4in]{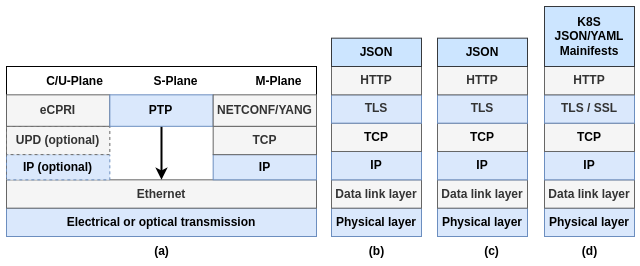}
\caption{Protocol stacks for O-RAN interfaces,  showcasing (a) Open Fronthaul data plane encapsulation, (b) A1 interface, (c) O1 interface, and (d) O2 interface.}
\label{fig:A1O1O2FH}
\vspace{-5mm} 
\end{figure}
\subsection{\textbf{Open Fronthaul Interface}}
In this section we explore how the Open FH Interface (OFHI) can be made compatible with the O-LoRaWAN DU-RU communication using the Enhanced Common Public Radio Interface (eCPRI) protocol \cite{ecpri2019specification}. Note that in the traditional LoRaWAN architecture the gateway functionalities are mostly monolithic and cannot be split. 

The OFHI supports different message types across User, Control, Management, and Synchronization planes. In its more simple deployment builds on Ethernet for the physical communication between the RU and DU components, as illustrated in Fig.\ref{fig:A1O1O2FH}(a). We place particular emphasis on the Control and User planes as LoRaWAN gateways, analogously to the O-RAN architecture, rely on IP-based communication for the management and synchronization planes, which facilitates the integration of protocols like NETCONF~\cite{rfc6241} or PTP \cite{rfc8173}. 
 
In contrast the UP and CP in the O-RAN architecture, rely on the eCPRI protocol. In O-LoRaWAN this interface needs to be adapted from the aforementioned monolithic approach.
\begin{figure}[t]
\centering
\includegraphics[width=3.4in]{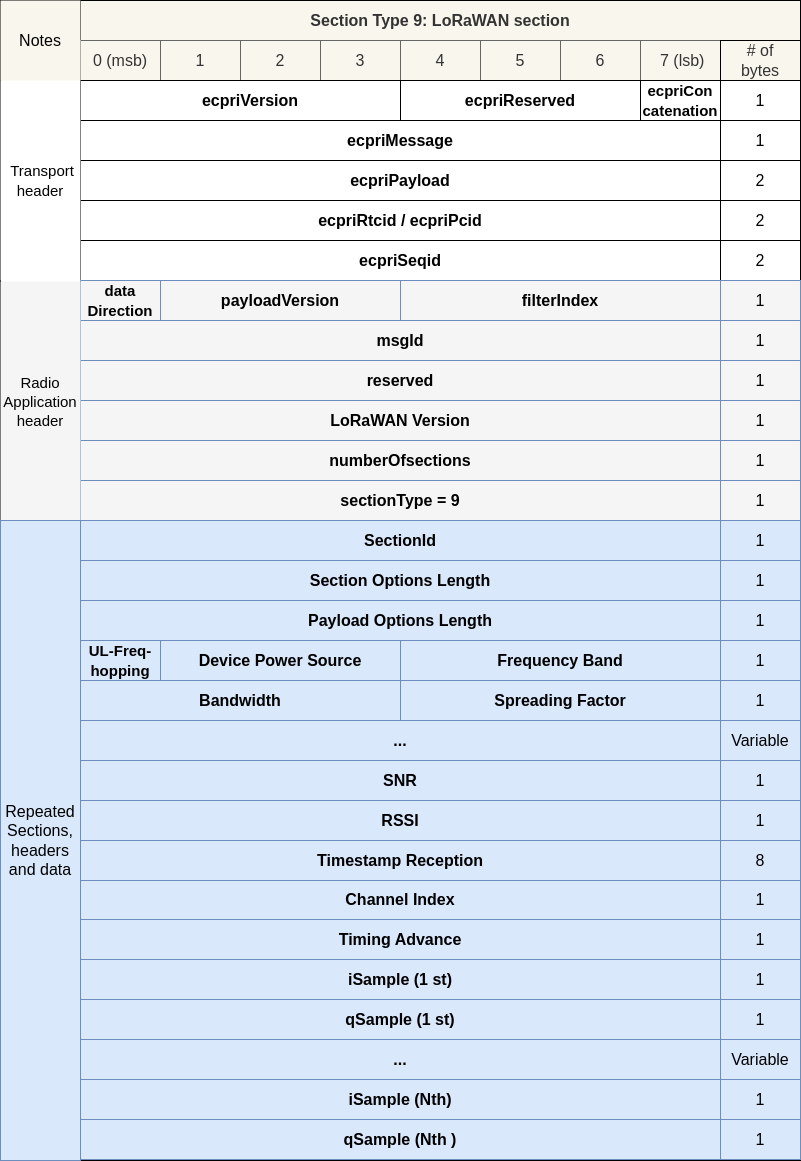}
\caption{Frame structure for LoRaWAN section type utilized for UL in the O-LoRaWAN.}
\label{fig:section_type_o_lorawan_dl_ul}
\vspace{-5mm} 
\end{figure} 

The eCPRI protocol encapsulates information through a two layer structured messages over Ethernet (or IP if systems are accessible trough IP networks). The first layer consists of an eCPRI common header, including corresponding fields used to indicate the message type and control the eCPRI flow. The second layer is an application layer including necessary fields for the RU data transportation, control and synchronization. Within the application layer, a "section" defines the information and data elements of the User Plane to be transferred or received from the radio.

The type of the transported message is identified by an 8-bit Section Type field in the common header. Section Types are defined in Table 7.3.1-37 of the specification~\cite{O_RAN_WG4_2023}, and determine the functions that can be transported between the O-DU and O-RU. The current specification is tailored to the physical and MAC layer of cellular systems. For example, considering information is identified by the resource block which is not adaptable to the physical and MAC characteristics of LoRaWAN.  

For that reason we propose a particular extension of the eCPRI Section Types so LoRaWAN can be properly encapsulated. Particularly we propose a new Section Type message, referred to as LoRaWAN Section with id 0x09, using one of the reserved values in the 0x09-0xFF range. This sort of extension can be applied to other protocols, adapting the particular application/technology content of the eCPRI transport to the target radio technology. 

Fig.\ref{fig:section_type_o_lorawan_dl_ul}, presents the structure of the LoRaWAN Section which presents the information on the headers and segments. This structure adheres to the OFHI frame format. While the radio application header and repeated sections provide detailed coding unique to section types, the transport header contains standard coding elements like the eCPRI version and message type.

In addition, Table \ref{tab:O-LoRaWAN_Section_Type} lists the parameters that need to be incorporated in the LoRaWAN Section in order to ensure the RU is properly configured and the DU has the proper physical layer information to process the frames. The parameters are listed here due to space limitations in Fig. \ref{fig:section_type_o_lorawan_dl_ul}. 

\begin{table}[h]
\renewcommand{\arraystretch}{2}
\begin{center}
\begin{adjustbox}{width=.5\textwidth} 
\begin{tabular}{ p{4cm} p{1cm} p{4cm} p{1cm} p{1cm} }

\hline
\rowcolor{blue!15}
\textbf{Attribute} & \textbf{Direction} & \textbf{Possible Values} & \textbf{Requirement} & \textbf{Length (bits)} \\

iSample & UL & In-phase samples & Optional & Variable \\

qSample & UL & Quadrature samples & Optional & Variable \\

Demodulation Information & UL & Chirp demodulation specifics & Required & Variable \\

Uplink Channel SNR & UL & Decibel values & Required & 8 \\

Battery Status of Device & UL & Percentage or voltage level & Optional & 8 \\

Uplink Frequency Hopping & UL & Enabled/Disabled & Optional & 1 \\

Timestamp Reception &  UL & Exact moment when the message was received, usually in UTC & Required & 64  \\

Uplink Channel RSSI & UL & Decibel values relative to 1mw & Required & 8 \\

Uplink Channel Utilization & UL & Percentage of time, channel & Optional & 8 \\

Device Power Source & UL & Battery, USB, Solar, Mains & Optional & 3 \\

Timing Advance & UL & Timing adjustment values & Required & 8 \\

Receive Window Configuration & UL/DL & Time slots post DL transmission & Optional & 8 \\

Channel Plan Configuration & UL/DL & Default, custom, regional & Optional & 4 \\

Frequency Band & UL/DL & Sub-GHz ISM bands & Optional & 4 \\

Spreading Factor & UL/DL & SF7 to SF12 & Required & 4 \\

End-device Firmware Version & UL/DL & Semantic versioning format & Optional & Variable \\

Preamble Length & UL/DL & Number of preamble symbols & Optional & 8 \\

Antenna Selection & UL/DL & Antenna indices or identifiers & Optional & Variable \\

Channel Index & UL/DL & Channel numbers based on plans & Optional & 8 \\

Bandwidth & UL/DL & 125 kHz, 250 kHz, or 500 kHz & Required & 4 \\

LoRaWAN Version & UL/DL & Version numbers like 1.0.x, etc. & Required & 8 \\

Data Direction & UL/DL & Uplink (UL), Downlink (DL) & Required & 1 \\

Payload Version & UL/DL & Version of payload format & Required & 3 \\

Filter Index & UL/DL & Index used for processing data & Optional & 4 \\

Section ID & UL/DL & Unique identifier for the section & Required & 8 \\

Section Options Length & UL/DL & Length of the section options field & Required & 8 \\

Device Address & DL & 32-bit device address & Required & 32 \\

Downlink Payload Content & DL & Sensor data, commands, acknowledgments. & Required & Variable \\

Frequency Hopping Pattern & DL & Specific channel sequences & Optional & 8 \\

Transmission power level & DL & Ranges typically from 2 dBm to 20 dBm & Required & 8 \\

Transmission Slot & DL & Absolute or relative time slots & Required & 64 \\

RX Window Configuration & DL & Timing, duration, and frequency & Optional & 16 \\

Device Class Type & DL & A, B, C & Optional & 2 \\

Energy Efficiency Considerations & DL & Power-saving modes, wake-up & Optional & 8 \\

Network Synchronization & DL & GPS timing, NTP & Optional & Variable \\

Traffic Prioritization & DL & Priority levels & Optional & Variable \\

Beacon Broadcasting & DL & Beacon intervals and payloads & Optional & Variable \\

\hline
\end{tabular}
\end{adjustbox}
\end{center}
\caption{O-LoRaWAN Section Type Specifications Attributes within the Open-FH Interface Framework}
\label{tab:O-LoRaWAN_Section_Type}
\vspace{-5mm}
\end{table}

\subsection{\textbf{E2 interface}}
\begin{figure}[ht]
\centering
\includegraphics[width=3.4in]{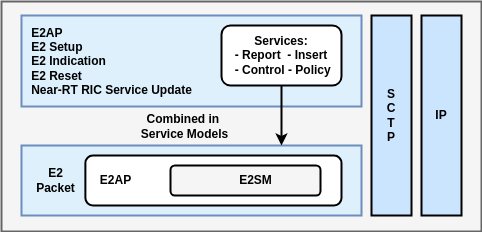}
\caption{E2 packet format \cite{O_RAN_WG3_E2GAP_2023}.}
\label{fig:E2}
\end{figure} 

The E2 interface is designed to enable the communication between the near-RT RIC and other network elements like the O-LoRaWAN-DU or O-LoRaWAN-NS.
In O-LoRaWAN, the interface must serve for the performance control and operation of the network via the decisions taken by the xAPPs, the collection of performance metrics from the different O-LoRaWAN components.
It runs over the Stream Control Transmission Protocol (SCTP) on IP. At the application layer, information and actions are encapsulated using two protocols: E2 Application Protocol (E2AP) and E2 Service Model (E2SM), refer to Fig.~\ref{fig:E2}. 
E2AP manages connections and transport of the information, while E2SM focuses on encoding message content with network performance metrics and other service-related information. This dual-protocol approach allows for the encapsulation of the actions (notifications, requests) and the required information for the near-RT RIC to feed xApps and make informed decisions based on real-time network analytics.
The O-LoRaWAN E2 key functions include:\\
\textbf{Control Messaging Support:} Enables the near-RT RIC to send control messages, including network configuration updates and downlink scheduling instructions, to O-LoRaWAN components.\\
\textbf{Network Insights and Performance Monitoring:} Facilitates the exchange of performance metrics and traffic data, supporting network analysis and decision-making for upgrades.\\
\textbf{Resource Management Assistance:} Aids in resource allocation by providing the near-RT RIC with real-time analytics and predictions.\\
\textbf{Dynamic Configuration:} Supports networking tasks such as device registration and network settings adjustments based on live information.\\
\textbf{Anomaly Detection and Reporting:} Ensures network integrity by facilitating timely reporting and correction of network issues.

The E2 interface uses the ASN.1 standard for data structure description, enhancing interoperability in the multi-vendor O-LoRaWAN environment. This standardization is pivotal for the growth of the LoRaWAN back-end and server infrastructure, enabling vendors to build their components with interoperability by design.

\subsection{\textbf{A1 interface}}
The A1 interface provides the communication standard between the SMO framework and the Near-RT RIC. It ensures the transition of policies to the Near-RT RIC. Subsequently, the Near-RT RIC turns these policies into specific configurations for network entities running the O-LoRaWAN. Moreover, it supports applications and services through its policy management layer (A1-P) and enrichment layer (A1-EI) functions. The A1 interface is key to defining how devices join the network and managing power saving schemes. It also facilitates the development of differentiated services, for example providing priorities to certain devices when required among other applications.

As depicted in Fig.\ref{fig:A1O1O2FH}(b), showcasing the A1 protocol stack, the A1 interface uses HTTP/HTTPS as application layer protocols. It enables security by applying TLS encryption over the data sent between the Non-RT RIC and the Near-RT RIC. Data across the interface is serialized in JSON and transmitted via a REST API. This architecture supports efficient Create, Read, Update, and Delete (CRUD) operations on network resources; subsequently, operation scales up and becomes interoperable in network policy and function management.

In O-LoRaWAN through the A1 interface, high-level policies are applied to both user and control planes, affecting routing, data prioritization, channel access, and device management. These policies guide the O-LoRaWAN Near-RT RIC, impacting gateway operations and the management of O-LoRaWAN-DU and O-LoRaWAN-RU elements. The A1 interface ensures network performance and efficiency by managing O-LoRaWAN-DU/RU selection, load balancing, and transmission power adjustment, optimizing resource utilization.
\subsection{\textbf{O2 interface}}
In O-LoRaWAN, the O2 interface is employed in resource management operations of the cloud or virtualized infrastructure that supports the O-LoRaWAN network. It brings the SMO framework and cloud resources together in order to make sure the network is provided with sufficient computational and storage resources. It enables Cloud service’s management, supporting its customization to the requirements of the O-LoRaWAN network. Table \ref{tab:o1_o2_functionalities} defines some functions of the O2 interface on the O-LoRaWAN network, detailed in the right column.

\subsection{\textbf{O1 interface}}
The O1 interface connects O-LoRaWAN components with the management entities in the SMO framework. The SMO framework in the legacy LoRaWAN is not standardized and NS and AS components are not managed as part of a virtualized managed framework (VNF), thus automation of infrastructure operation is a non-standardized featured developed by the different operators (e.g Actility, TTI, etc..).

O1 and O2 are an opportunity to standardize the infrastructure operation, management and deployment. O1 builds on widely used standards for network operation (refer to  Fig.~\ref{fig:A1O1O2FH}(c)), utilizing NETCONF and YANG data model transported on secure TLS/TCP/IP links. The integration of LoRaWAN infrastructure management into O-LoRaWAN will 
enable the network to improve its operations, especially in fault management, scaling,  simplified configuration, AI model update or provision in xApps or rApps, and security provision. Table \ref{tab:o1_o2_functionalities} defines some functions of the O1 interface on the O-LoRaWAN network.

\begin{table}[t]
\renewcommand{\arraystretch}{2}
\begin{center}
\begin{adjustbox}{width=.48\textwidth} 
\begin{tabular}{ p{5cm} p{5cm} }
\hline
\rowcolor{blue!15}
\textbf{Functionalities of the O1 in O-LoRaWAN} & \textbf{Functionalities of the O2 in O-LoRaWAN} \\
\hline

\textbf{Comprehensive Configuration Management:} Seamlessly updates software, security patches, and optimizes network parameters across O-LoRaWAN-NS, O-LoRaWAN-DU, Near-RT RIC and O-LoRaWAN-RU to ensure peak performance and reliability. & 
\textbf{Resource Management:} Allocates, scales, and dynamically adjusts virtual resources such as processing power and memory for O-LoRaWAN entities like the O-LoRaWAN-NS, and O-LoRaWAN-DU. This is essential for efficiently managing network traffic and optimizing resource use. \\

\textbf{Policy Distribution and Dynamic Configuration:} Distributes operational policies, including data routing protocols, load balancing, and device connectivity policies, enabling dynamic network adjustments based on real-time analysis. & 
\textbf{Service Lifecycle:} Orchestrates the deployment, configuration, and integration of O-LoRaWAN components. It manages the instantiation and termination of services, ensuring they scale elastically to meet network demands. \\

\textbf{Fault Management and Issue Resolution:} Identifies and resolves operational issues, observing the network components and traffic status for both user and control plane elements. &
\textbf{Maintenance and Security:} Automates updates, rollbacks, and security enhancements for O-LoRaWAN components, keeping the network up-to-date and secure. \\

\textbf{Performance Optimization:} Gathers key performance indicators (KPIs) such as signal strength and uptime to assess and improve functionality, optimizing network efficiency. & 
\textbf{Instantiation and Termination:} Manages the instantiation of virtual O-LoRaWAN entities and other O-LoRaWAN components instances and their termination when they are no longer needed, ensuring efficient use of cloud infrastructure. \\

\textbf{Lifecycle Management:} Oversees the lifecycle of network elements from deployment to decommissioning, assessing proper operations. & 
\textbf{Data Management:} Manages cloud storage for data collected from the O-LoRaWAN network. \\

\hline
\end{tabular}
 \end{adjustbox}
\end{center}
\caption{Functionalities of the O1 and O2 Interfaces in O-LoRaWAN}
\label{tab:o1_o2_functionalities}
\vspace{-5mm}
\end{table}


\section{\textbf{Conclusion}}
\label{sec:conclusions}

In this article, we proposed transforming the current architecture of LoRaWAN networks into a modular architecture that follows the O-RAN model. The vision is to standardize network architectures so that they can pivot to a network control, orchestration, and management model that allows for the reuse of functions. This standardization will enable the development and extension of network functions based on micro-applications, that could even be reused for different RANs. LoRaWAN is an interesting candidate because it shares some similarities to cellular networks, the main focus of O-RAN.

We propose O-LoRaWAN, the adaptation of the LoRaWAN architecture to the O-RAN framework. In the article we describe how components can be adapted and how the current LoRaWAN functions can be split. There are three major adaptation paths: first, the splitting of the LoRa gateway functionalities, handling them in two main components, the Radio Unit and the Distributed Unit. A second adaptation, is on the network server functions and the integration of RIC concepts to support data exploitation in the form of \textit{Apps}, a model that will enable wider development of the LoRaWAN functions. And third, the standardization of the operations, orchestration and management of network components and underlying infrastructure. The SMO is not standardized in the current LoRaWAN deployments.

The article analyzes the main O-RAN interfaces and suggests possible adaptations in the supporting protocols to ensure compatibility between components. In this adaptation, we identify ways forward for the growth of O-RAN towards supporting different RANs. This involves designing protocols that can be extended via Information Elements.

%
\IEEEpeerreviewmaketitle


%

\section*{Acknowledgment}
This project is co-financed by the European Union’s Horizon 2020 program under Marie Skłodowska-Curie grant agreement No 813999. This project is also co-funded by the Spanish Ministry of Science, Innovation, and Universities under RF-VOLUTION project (PID2021-122247OB-I00) and the 2021 SGR 00174 From Generalitat de Catalunya. Prof. Xavier Vilajosana is sponsored by ICREA Academia Grant. This research is also supported by the 6G-OASIS-Secure project (UNICO-5G I+D) TSI-063000-2021-77.

\ifCLASSOPTIONcaptionsoff
  \newpage
\fi



%
\bibliography{references}

\begin{thebibliography}{10}

\bibitem{adelantado2017understanding}
F.~Adelantado, X.~Vilajosana, P.~Tuset-Peiro, B.~Martinez, J.~Melia-Segui, and T.~Watteyne, ``Understanding the limits of lorawan,'' {\em IEEE Communications magazine}, vol.~55, no.~9, pp.~34--40, 2017.

\bibitem{9422331}
G.~Boquet, P.~Tuset-Peiró, F.~Adelantado, T.~Watteyne, and X.~Vilajosana, ``Lr-fhss: Overview and performance analysis,'' {\em IEEE Communications Magazine}, vol.~59, no.~3, pp.~30--36, 2021.

\bibitem{3GPP2017}
{3rd Generation Partnership Project (3GPP)}, ``Study on new radio access technology: Radio access architecture and interfaces, version 14.0.0,'' Technical Report TR 38.801, 3rd Generation Partnership Project (3GPP), Sophia Antipolis, France, Apr. 2017.
\newblock [Online]. Available: \url{http://www.3gpp.org/DynaReport/38801.html}.

\bibitem{O_RAN_WG1_2023}
{O-RAN Alliance}, ``{O-RAN Architecture Description -- v10.00.00},'' tech. spec., O-RAN Work Group 1 (Use Cases and Overall Architecture), December 2023.
\newblock Technical Specification.

\bibitem{O_RAN_WG2_NONRTRIC_ARCH_2023}
{O-RAN Alliance}, ``{Non-RT RIC Architecture -- v04.00.00},'' tech. spec., {O-RAN Work Group 2 (Non-RT RIC and A1 interface)}, 2023.

\bibitem{O_RAN_WG2_A1GAP_2023}
{O-RAN Alliance}, ``{A1 interface: General Aspects and Principles -- v03.01},'' tech. spec., {O-RAN Work Group 2 (Non-RT RIC and A1 interface)}, 2023.

\bibitem{O_RAN_WG2_NONRTRIC_FUNARCH_2021}
{O-RAN Alliance}, ``{Non-RT RIC: Functional Architecture -- v01.01},'' technical report, {O-RAN Work Group 2 (Non-RT RIC and A1 Interface)}, 2021.

\bibitem{O_RAN_WG3_RICARCH_2023}
{O-RAN Alliance}, ``{Near-RT RIC Architecture -- v05.00.00},'' tech. spec., {O-RAN Work Group 3 (Near-Real-time RAN Intelligent Controller and E2 Interface)}, 2023.

\bibitem{O_RAN_WG4_2023}
{O-RAN Alliance}, ``Control, user and synchronization plane specification -- v13.00.00,'' tech. spec., O-RAN Work Group 4 (Open Fronthaul Interfaces WG), 2023.

\bibitem{O_RAN_WG6_CADS_2023}
{O-RAN Alliance}, ``{Cloud Architecture and Deployment Scenarios for O-RAN Virtualized RAN -- v05.00.00},'' technical report, {O-RAN Working Group 6 (Cloudification and Orchestration)}, 2023.

\bibitem{O_RAN_WG6_O2GAP_2022}
{O-RAN Alliance}, ``{O-RAN O2 Interface General Aspects and Principles -- v01.02},'' technical report, {O-RAN Working Group 6 (Cloudification and Orchestration)}, 2022.

\bibitem{O_RAN_WG3_E2GAP_2023}
{O-RAN Alliance}, ``{E2 General Aspects and Principles (E2GAP) -- v04.01},'' technical specification, {O-RAN Work Group 3 (Near-RT RIC and E2 Interface)}, 2023.

\bibitem{O_RAN_WG1_O1_2021}
{O-RAN Alliance}, ``O-ran operations and maintenance interface specification -- v04.00,'' technical specification, O-RAN Work Group 1 (O1 Interface Specification), 2021.

\bibitem{polese2023understanding}
M.~Polese, L.~Bonati, S.~D’oro, S.~Basagni, and T.~Melodia, ``Understanding o-ran: Architecture, interfaces, algorithms, security, and research challenges,'' {\em IEEE Communications Surveys \& Tutorials}, 2023.

\bibitem{LoRaWAN2017Specification}
{LoRa Alliance}, ``{LoRaWAN 1.1 Specification, Oct. 2017}.'' \url{http://lora-alliance.org/lorawan-for-developers}.

\bibitem{ecpri2019specification}
{Ericsson AB et al.}, ``{eCPRI Specification V2.0},'' {\em Common Public Radio Interface: eCPRI Interface Specification}, vol.~2, pp.~1--, 2019.
\newblock Available: http://www.cpri.info/spec.html.

\bibitem{rfc6241}
R.~Enns, M.~Björklund, A.~Bierman, and J.~Schönwälder, ``{Network Configuration Protocol (NETCONF)}.'' RFC 6241, June 2011.

\bibitem{rfc8173}
V.~Shankarkumar, L.~Montini, T.~Frost, and G.~Dowd, ``{Precision Time Protocol Version 2 (PTPv2) Management Information Base}.'' RFC 8173, June 2017.

\end{thebibliography}
\bibliographystyle{ieeetr}

%

\begin{IEEEbiography}[{\includegraphics[width=1in,height=1.25in]{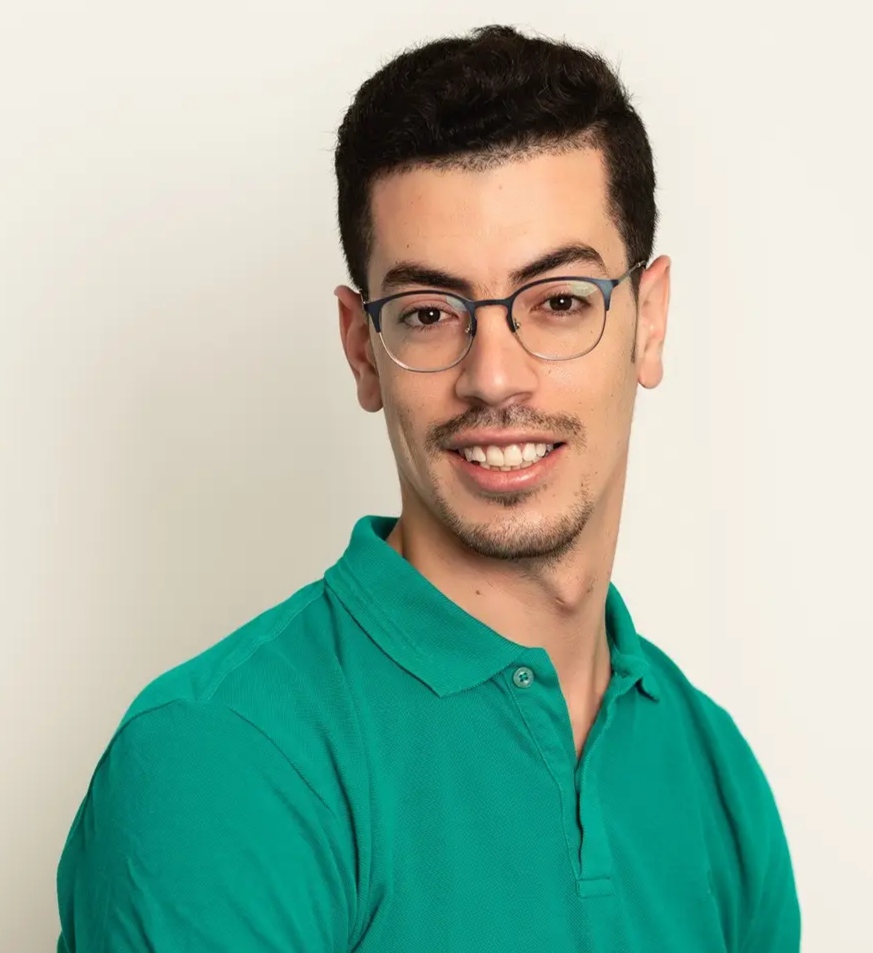}}]{Sobhi Alfayoumi} 
Holding a B.Sc. in Communications and Computer Engineering from Al-Azhar University-Gaza and an M.Sc. from the University of Padova, Italy, he transitioned from a ML engineer role in Italy to research at Spain's Imdea Network Institution. He is currently pursuing an industrial PhD with Worldsensing Company and the Universitat Oberta de Catalunya (UOC), also serving as a Senior Research at UOC's WiNE Research Group.
\vspace{-3em}
\end{IEEEbiography}

\begin{IEEEbiography}[{\includegraphics[width=1in,height=1.25in ]{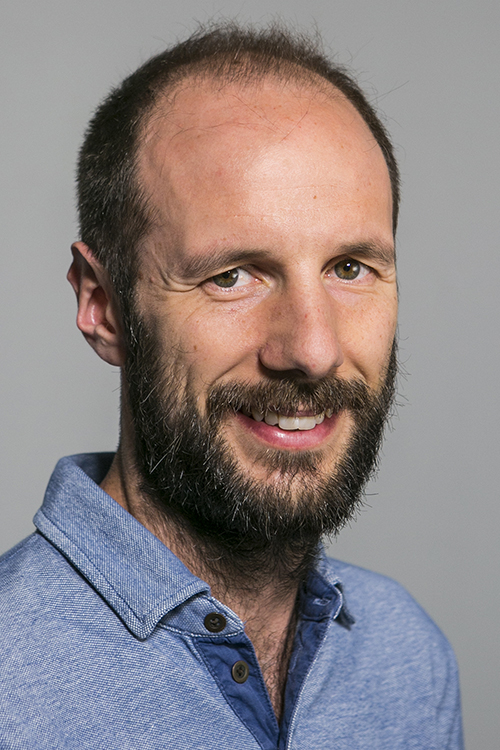}}]{Joan Melià-Seguí}(Senior Member, IEEE) received the B.Sc. and M.Sc. degrees in telecommunications engineering from the Universitat Politècnica de Catalunya and the Ph.D. degree from the Universitat Oberta de Catalunya (UOC). He has been a Researcher at the Universitat Pompeu Fabra, Visiting Researcher at the Palo Alto Research Centre (Xerox PARC), and Fulbright Visiting Scholar at the Massachusetts Institute of Technology. Currently, he is an Associate Professor at UOC, specializing in sustainability and low-cost RF sensing. 
\vspace{-3em}
\end{IEEEbiography}

\begin{IEEEbiography}[{\includegraphics[width=1in,height=1.25in]{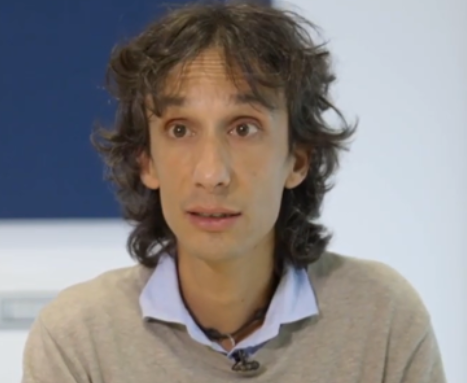}}]{Xavier Vilajosana}
(Senior Member, IEEE) received the B.Sc. and M.Sc.
degrees in computer science from the Universitat Politècnica de Catalunya,
Barcelona, Spain, in 2004 and 2006, respectively, and the Ph.D. degree in
computer science from the Universitat Oberta de Catalunya (UOC), Barcelona,
in 2009.
He has been a Researcher with Orange Labs, France, HP,
Barcelona, and UC Berkeley. He is an ICREA Academia
Professor with UOC.
\end{IEEEbiography}




\end{document}